\documentclass{desyproc}
\usepackage{graphicx}
\begin{document}

\title{\vspace{-3cm}{\small \hfill{{DESY 12-036}}}\\[1.8cm]
Cold Dark Matter from the Hidden Sector}


\author{{\slshape Paola Arias$^{1,2}$\\[1ex]
$^1$Deutsches Elektronen-Synchrotron (DESY), Hamburg, Germany\\
$^2$Facultad de F\'isica, Pontificia Universidad Cat\'olica de Chile, Casilla 306, Santiago 22, Chile
}}
\contribID{arias\_paola}

\desyproc{DESY-PROC-2011-04}
\acronym{Patras 2011} 

\maketitle

\begin{abstract}
Weakly interacting slim particles (WISPs) such as hidden photons (HP) and axion-like particles (ALPs) have been proposed as cold dark matter candidates. They might be produced non-thermally via the misalignment mechanism, similarly to cold axions. In this talk we review the main processes of thermalisation of HP and we compute the parameter space that may survive as cold dark matter population until today. Our findings are quite encouraging for experimental searches in the laboratory in the near future.

\end{abstract}

\section{Introduction}

WISPy cold dark matter (CDM) is not so surprising  after  all, since WISPs fulfill one of the prime conditions that one may ask to cold dark matter: a very weak coupling to Standard Model (SM) particles. A good example is the axion, produced non-thermally by the misalignment mechanism \cite{Preskill:1982cy,Abbott:1982af,Dine:1982ah}. However, it has been proposed recently \cite{Nelson:2011sf} that also cold dark matter comprised of light hidden Abelian gauge bosons - known as paraphotons or hidden photon (HP)  - might  be produced by this mechanism. As a matter of fact,  the misalignment mechanism is quite general and can generate CDM out of any light boson field that satisfy some general conditions \cite{wispycdm}. In the second part of the manuscript we discuss detection techniques and present constraints.

\section{WISPy cold dark matter and the misalignment mechanism}

The misalignment mechanism was originally conceived for the non-thermal production of axions. During the QCD phase transition era the initial angle of the axion field starts to realign with the true QCD vacuum. The energy in the field then behaves like dark matter. Such a mechanism can also produce a population of a different boson field provided its mass is smaller than the characteristic time of the universe ($t_0 \sim H$), $m \, t_0 \ll 0$. When $m \, t_0 \sim 1 $ is reached the field is able to respond to the existence of its mass, readjusting the state to a different minimum of the potential. Once the minimum is reached, certain overshoot is expected, causing the oscillation of the field around it.  The smaller the mass of the field, the later it will begin to oscillate, allowing for bigger dark matter abundances. Therefore, the misalignment mechanism could naturally give rise to a population of WISPy CDM, provided it does not thermalise until today.

For instance, let us  consider the case of the hidden photon, which occur naturally in many SM extensions based on field and string theory \cite{Goodsell:2010ie}. Assuming the universe underwent a period of inflation at a value of the Hubble expansion parameter ($H$) larger that the HP mass, $ m_{\gamma'} \ll H$, where $m_{\gamma'}$ is the HP mass, the equation of motion for the spatial components in an expanding Universe reads~\cite{Nelson:2011sf},
\begin{equation}
\ddot{B}^{i}+3H\dot{B}^{i}+m^{2}_{\gamma^{\prime}}B^{i}=0 .
\end{equation}
Details can be found in \cite{wispycdm}.
The equation undergoes two different regimes: $H\gg m_{\gamma'}$, the field gets essentially frozen and cannot respond to the mass term.
At a later time, $t_{\rm initial}$, at which $H(t_{\rm initial})\sim m_{\gamma'}$ the damping becomes undercritical and the field can roll down the potential and start to oscillate. The residual damping due to the expansion factor reduces the amplitude of the oscillations, such that the energy density of the HP field scales as $ \frac{1}{2}m^{2}_{\gamma^{\prime}}(B^{i})^2\sim a^{-3},$
which is exactly the behavior expected for non-relativistic matter.
The dark matter density of the HP field today depends on the initial {\it misalignment}, $B^{i}_{\rm initial}$, of the field therefore to achieve the correct dark matter density requires a fine-tuning.

\section{Cosmological constraints on hidden photons}
\begin{figure}[t]
\begin{minipage}[b]{0.5\linewidth}
\centering
\includegraphics[scale=0.78]{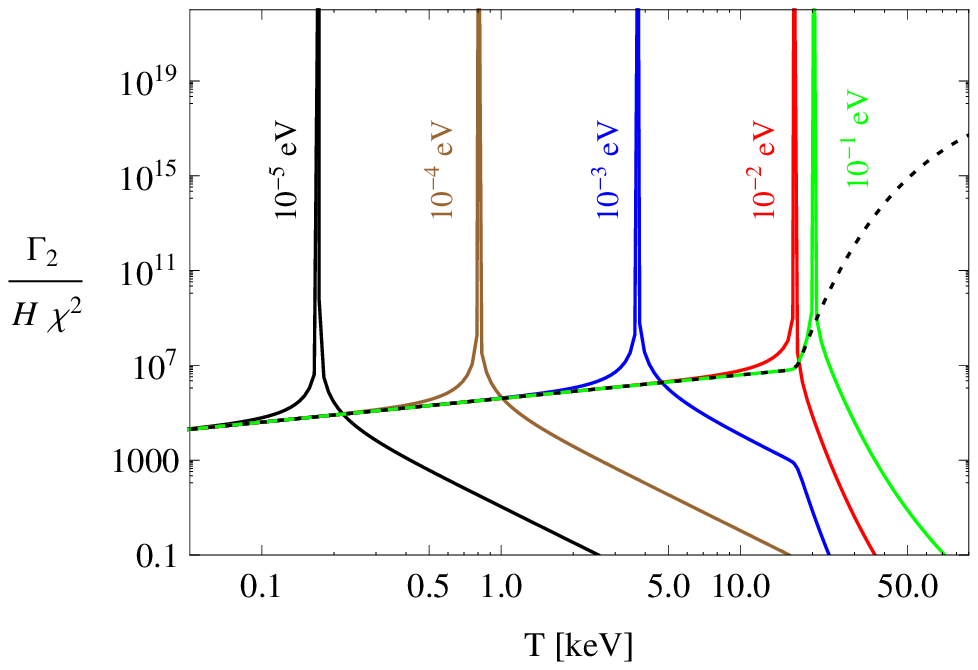}
\caption{\footnotesize{ $\Gamma/\left(\chi^2 H\right)$ for different hidden photon masses. The effects near the resonance clearly dominates. The corresponding ÒapproximationÓ from reference \cite{Nelson:2011sf} is shown as a dotted line.}}
\label{fig:figure1}
\end{minipage}
\hspace{0.5cm}
\begin{minipage}[b]{0.5\linewidth}
\centering
\includegraphics[scale=0.5]{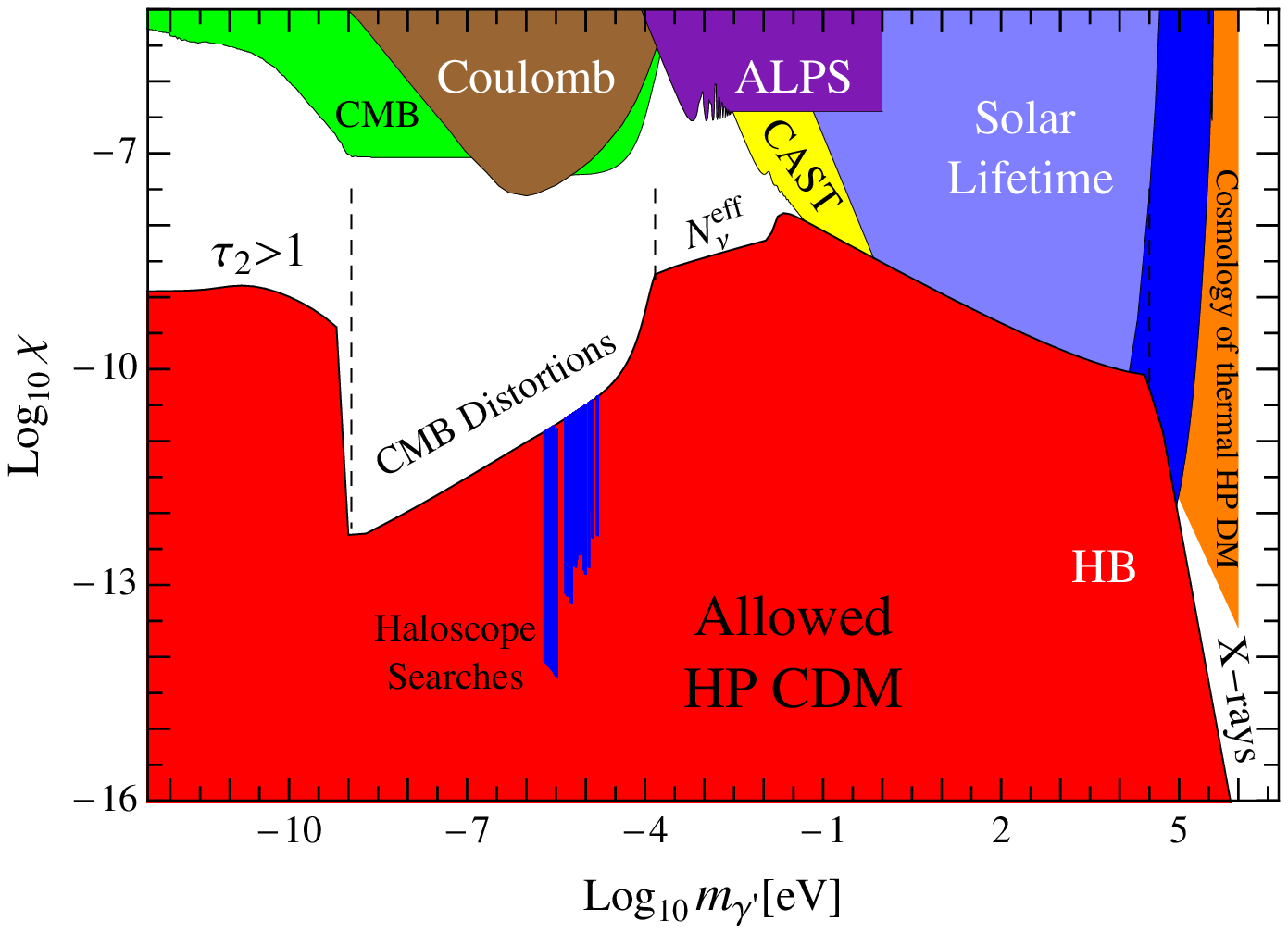}
\caption{\footnotesize{The region in red corresponds to where the HP CDM condensate does not thermalise. The blue region corresponds to the the constraints obtained from direct dark matter searches. For details of the figure see  \cite{wispycdm}.}}
\label{fig:figure2}
\end{minipage}
\end{figure}

In order to ensure that the boson field produced by the misalignment mechanism survives until today one needs to check whether there is some process that could thermalise it.
Is generally assumed that the SM particles are uncharged under the HP and the only interaction between the visible and hidden sector is through kinetic mixing between photons and HPs \cite{Holdom:1985ag}
\begin{equation}
\label{HPlagrangian}
\mathcal L= -\frac{1}4F_{\mu\nu}F^{\mu\nu}-\frac{1}4B_{\mu\nu}B^{\mu\nu}+\frac{m_{\gamma'}^2}2 B_\mu B^\mu-\frac{\chi}2 F_{\mu\nu} B^{\mu\nu}+ J^\mu A_\mu, \end{equation}
where the $A_\mu$ and $B_\mu$ correspond to photon and hidden photon field, respectively. The mixing between both is parametrised by the kinetic mixing term $\chi$ and we have included the coupling between electrically charge matter described by the current $J_\mu$, and photons. Nevertheless, once a change of basis of the type $A_\mu\rightarrow \tilde A_\mu -\chi \tilde B_\mu$, $B_\mu\rightarrow \tilde B_\mu$ to remove the kinetic mixing, one can see that ordinary matter also couples to the massive hidden photon state.  Thus, the HP couples to charged matter with a strength $\chi$.  However, in the early universe the plasma effects modify the effective kinetic mixing dramatically (details can be seen in \cite{Redondo:2008aa,Jaeckel:2008fi, Mirizzi:2009iz,Redondo:2008ec}).  The effective mixing parameter shows a resonance at the effective photon mass $m_{\gamma}$,
\begin{equation} \chi^2 \rightarrow\chi_{\rm eff}^2= \frac{\chi^2 m_{\gamma'}^4}{\left(m_\gamma^2-m_{\gamma'}^2\right)^2+\left(\omega \Gamma_0\right)^2},\label{chieff}
\end{equation}
where $m_{\gamma}$ is the photon mass in a plasma and $\omega \Gamma_0$ is a damping coefficient associated with absorption part of the refractive index of the medium. Examining the possible processes that could thermalise  the condensate, the Compton evaporation $\gamma'+e\rightarrow \gamma +e$ may be a dominant process, as discussed in ~\cite{Nelson:2011sf}.  A close estimate of the condition $\Gamma/H < 1$, where $\Gamma $ is the interaction rate of Compton evaporation,
\begin{equation} 
\Gamma= \int d^3p \, \sigma(p)\, v(p)\, f_e(p), 
\label{gamma}
\end{equation}
and the Hubble expansion rate,  
$
H= 1.66\, g_*^{1/2} {T^2}/{M_P},
$
shows that an important contribution to the ratio $\Gamma/H$ comes from the resonance condition $m_{\gamma}\sim m_{\gamma'}$, associated with a resonance temperature, $T_{\rm res}$. The result, for different HP masses, is plotted in Fig.~\ref{fig:figure1}. As expected the decay rate is heavily enhanced at $T_{\rm res}$, corresponding to $m_{\gamma}=m_{\gamma'}$, and rapidly drops at higher temperatures. These effects have not been taking into account in Ref.~\cite{Nelson:2011sf} which neglected the plasma effects.
Performing a more careful computation at the resonance temperatures  we can obtain the maximal $\chi(m_{\gamma'})$ at which is possible to keep the condensate alive, displayed in Fig.~\ref{fig:figure2}.
This result is quite encouraging: hidden photons in this parameter range are not only predicted in popular string 
compactifications~\cite{Cicoli:2011yh}, but are also within reach of the next generation of pure laboratory experiments based on 
light-shining-through walls~\cite{Redondo:2010dp}. 

\section{Microwave cavity searches}
Microwave cavity experiments looking for relic axions could also be used to constrain and search for the hypothetical cold HP condensate that we have been discussing so far.
To start with, let us assume the energy density in the hidden photons is equal to the dark matter density, therefore,
$
\rho_{DM}={m^{2}_{\gamma^{\prime}}}/{2}|{\mathbf{B}}|^2.
$
To simplify things, we will consider two possible scenarios: the direction of the HP field is (essentially) unaffected by structure formation and all 
HPs point in the same direction (at least for a suitably big region of space). Or, the direction of HPs behave like a gas of particles with the vector pointing in random directions.
In  the first scenario the HP direction is characterized by a fixed vector $\hat{n}$, whereas in the second case we have to average
the final result over all directions for $\hat{n}$.
With this understood, let us write for the hidden photon field, 
$
\mathbf{B}(\mathbf{x})=\hat{\mathbf{n}} \frac{\sqrt{2\rho_{0}}}{{m_{\gamma'}}},
$
with $\rho_{0}$ the dark matter density on earth. 

In the microwave cavity,  the power emission is related to the energy stored inside, $U$, and the quality factor of the cavity, $Q$, as $\mathcal P_{\rm out}=\kappa \omega_0 U/Q$, where $\kappa$ is the coupling of the cavity to the detector. Evaluating at resonance ($\omega_0=m_{\gamma'}$) we find \cite{wispycdm}
\begin{equation}
\mathcal P_{\rm out}=\kappa \chi^2 m_{\gamma'} \rho\, Q\,  V\, \mathcal G,
\end{equation}
where the geometric factor 
$\mathcal G$ is defined as
\begin{equation} 
\mathcal G
= \frac{\left|\int dV{\mathbf{A}}^{*\rm cav}(\mathbf{x})\cdot \hat{\mathbf{n}}\right|^2}{V\, \int d^3\, \mathbf{x} |\mathbf{A}^{\rm cav}(\mathbf{x})|^2}.
\end{equation}
In a cavity ${\mathbf{E}}=\omega {\mathbf{A}}$. This geometry factor has exactly the same form as in the axion case (cf.~\cite{Asztalos:2001tf})
but with the direction $\hat{\mathbf{n}}$ replacing the direction of the external magnetic field $\mathbf{B}_{magnet}$ in the axion case. It is interesting that already now the negative searches for CDM axions of past microwave cavity experiments can be turned, via Eq. (5), into new exclusion regions for HPs, cf. the region "Haloscope Searches" in Fig. 2.
\section{Conclusions}
We have argued that the misalignment mechanism is a very powerful tool to produce a condensate of practically any light boson field in the early universe. In particular, in this letter we have focused on hidden photons. Our results show that the parameter space where the CDM population may have survived until today is accessible to future laboratory experiments, such as light-shining-through-walls experiments, helioscope experiments, and microwave cavity searches (haloscopes). All the above applies also to an axion-like particle (ALP) which can be considered to be a (pseudo-)scalar field $\phi$ enjoying a periodic shift symmetry - a common feature occurring in string compactifications. A generic ALP may have its mass arising from some hidden interactions, such as string instanton effects or explicit symmetry breaking. In some cases, the mass of the ALP can also vary in time, for instance due to thermal corrections. The results of these investigations will appear in a near future \cite{wispycdm}.

\section{Acknowledgments}
This talk is based on the arXiv paper arXiv:1201.5902 with Davide Cadamuro, Mark Goodsell, Joerg Jaeckel, Javier Redondo and Andreas Ringwald. P. A. acknowledges the valuable support of the Alexander von Humboldt Foundation.


\begin{footnotesize}

\end{footnotesize}


\end{document}